\documentclass[useAMS,usenatbib]{mn2e}

\usepackage{epsfig}
\usepackage{longtable}
\usepackage{times} 
\newcommand{\sw}{$Swift$}

\def \sw {{\em Swift}}

\def \apj {ApJ}

\def \apjs {ApJS}\def \aap {A\&A}
\def \mnras {MNRAS}

\title[  ]
{Finding a 61.0-day orbital period for the HMXB 4U~1036--56 with the Swift-BAT monitoring.}

\author[G.\ Cusumano et al.]{G.\ Cusumano$^{1}$,  A.\ Segreto $^{1}$ , V.\ La Parola$^{1}$, N.\ Masetti
$^{2}$, A.\ D'A\`i$^{3}$, G.\ Tagliaferri$^{4}$ \\
$^{1}$INAF - Istituto di Astrofisica Spaziale e Fisica Cosmica, Via U.\ La Malfa 153, I-90146 Palermo, Italy\\
$^{2}$INAF - Istituto di Astrofisica Spaziale e Fisica Cosmica di Bologna, via Gobetti 101, 40129, Bologna, Italy\\
$^{3}$Dipartimento di Fisica e Chimica, Universit\`a di Palermo, via Archirafi 36, 90123, Palermo, Italy\\
$^{4}$  INAF - Brera Astronomical Observatory, via Bianchi 46, 23807, Merate (LC), Italy
\\
}

\begin{document}

\date{}

\pagerange{\pageref{firstpage}--\pageref{lastpage}} \pubyear{2010}

\maketitle

\label{firstpage}

\begin{abstract}
Since November 2004, the Burst Alert Telescope on board Swift is producing a 
monitoring of the entire sky in the 15--150 keV band, recording the timing and 
spectral behavior of the detected sources.
Here we study the properties of the HMXB 4U~1036--56 using both the 
BAT survey data and those from a Swift-XRT observation. A folding
analysis performed on the BAT light curve of the first 100 months of survey unveils 
a periodic modulation with a period of 
$\sim 61.0$ days, tied to the presence in the BAT light curve of several 
intensity enhancements lasting $\sim 1/4$ of P$_0$. We explain this modulation 
as the orbital period of the binary system.
The position of 4U~1036--56 on the Corbet diagram, the derived semi-major 
orbit axis ($\simeq 180~R_{\odot}$), and the bulk of the source 
emission observed
in a limited portion of the orbit are consistent with a Be 
companion star. The broad band 0.2--150 keV spectrum is well 
modeled with a flat absorbed power law with a {\bf cut-off at $\sim 16$ keV}.
Finally, we explore  the possible association of 
4U 1036-56 with the $\gamma$-ray source AGL J1037-5808, finding that 
the BAT light curve does not show any correlation with 
the $\gamma$-ray outburst observed in November 2012.

\end{abstract}

\begin{keywords}
X-rays: binaries -- X-rays: individual: 4U~1036--56. 

\noindent
Facility: {\it Swift}

\end{keywords}


	\section{Introduction\label{intro}}

The identification of the orbital period of a binary system is an essential step 
for deriving  the geometry  of the system and, as a consequence,
for investigating the physical mechanism responsible for the spectral and 
temporal properties of the source. However, the discovery of such modulations   
may become a challenging task for very long periodicities, requiring long and 
continuous monitoring. For sources with strong absorption this monitoring
is efficient only in the hard X-ray regime where the circumstellar material is 
transparent to the emission, while it blocks most of the emission in the soft 
X-ray band. The long-term temporal monitoring carried on by the Burst Alert 
Telescope (BAT, \citealp{bat}) on board of the Swift observatory \citep{swift} 
has been fulfilling this task  (e.g. \citealp{ corbet1,
corbet2, corbet3, cusumano10, laparola10, dai11}).

\begin{figure*}
\begin{center}
\centerline{\includegraphics[width=12.cm,angle=0]{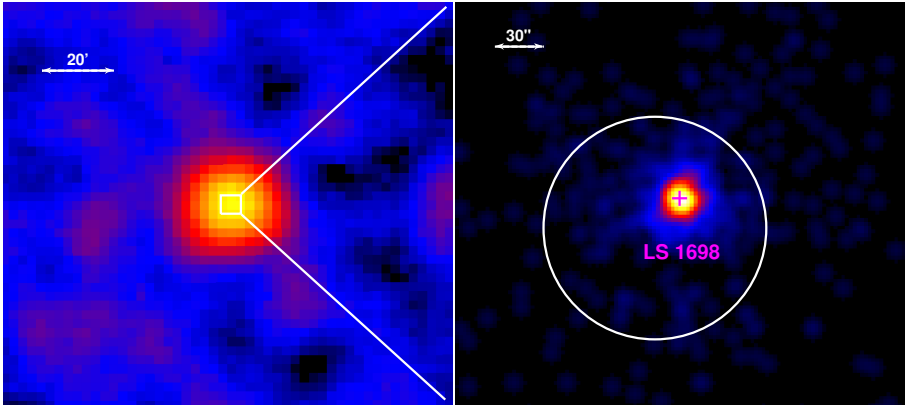}}
\caption[4U~1036--56 sky maps]{ Left panel: 15--60 keV significance map centered on 4U~1036--56.  
The box represents the sky area in the right panel.
Right panel: 0.2-10 keV XRT image; the circle represents the 95\% BAT error box of 1.16'.
The position of the optical counterpart (LS 1698) suggested by \citet{motch97} is marked 
with a cross.
                }
                \label{map} 
        \end{center}
        \end{figure*}

In this Letter we analyze the dataset collected by Swift on 4U~1036--56.
This source was first detected by UHURU \citep{forman78} and its transient 
behavior was first observed by Ariel V in 1974 \citep{warwick81}, that recorded 
an outburst episode with a flux increase of a factor of 2.4 higher than the 
average UHURU flux. Since then, 4U~1036--56 was observed by several X-ray 
telescopes, at different luminosity levels ranging between $10^{34}$ and 
$3\times 10^{35}$ erg s$^{-1}$ (see \citealp{lapalombara09} and references 
therein). RXTE data revealed a pulsed emission with a period of $860\pm 2$ s 
\citep{reig99}. Based on a pointed ROSAT PSPC observation \citet{motch97} 
identified the optical counterpart to be the B0 III-Ve star LS 1698, at a 
distance of $\sim$5 kpc. BAT detected a hard X-ray outburst from 4U 1036-56 
with a peak flux of $\sim 30$ mCrab on February 2012 \citep{atel3936}. This 
activated a Swift-XRT target-of-opportunity observation. The XRT spectrum was 
modeled with an absorbed flat power law with photon index $\sim 0.6$.
\citet{li12} report timing and spectral analysis of 4U~1036--56 with INTEGRAL
\citep{winkler03} and Swift data. They report an outburst detected in 
February 2007 with a significance $\sim 30$  and $\sim 10$ standard deviations 
in ISGRI and JEM-X, respectively. Their broad band spectrum can be described by 
an absorbed power-law (with photon index $\sim1.1$) modified by a cut-off at 
$\sim 26$ keV. Using the Swift-XRT observation, they confirm the pulsed 
modulation reported by \citet{reig99}. They also discuss the possibility 
that 4U~1036--56 is associated to the MeV source AGL~J1037-5708 (as suggested 
by \citealp{atel3059} upon the detection of the AGILE source). 
Using theoretical considerations based on the leptonic model for $\gamma$-ray 
emission, they do not rule out this possibility.\\
This Letter is organized as follows. Section 2 describes the BAT and XRT data 
reduction. Section 3 reports on the timing analysis. Sect. 4 describes the broad 
band spectral analysis.  In Sect.\ 5 we briefly discuss
our results. 

\begin{figure}[h]
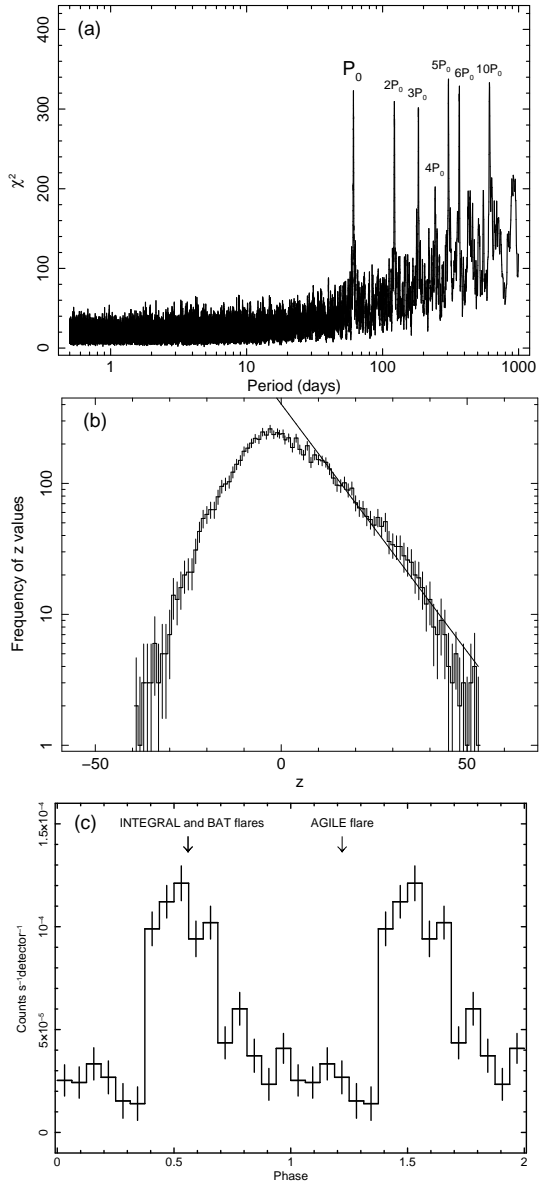

\begin{center}
\centerline{\includegraphics[width=5.2cm,angle=270]{4U1036-56_15-60_100mesin_r.ps}}
\centerline{\includegraphics[width=5.2cm,angle=270]{figura_histo_100mesi.ps}}
\centerline{\includegraphics[width=5.2cm,angle=270]{4U1036-56_15-60_xP0.ps}}

\caption[]{{\bf a}: Periodogram of \sw-BAT (15--60\,keV) data for 
4U~1036--56. 
{\bf b}: Distribution of the z values derived from the $\chi^2$ periodogram
(see Section~3). 
The positive tail beyond z=10 is modeled with an exponential function. 
{\bf c}: Light curve folded at a period P$_0$= 61.0\,days, with 16 phase 
bins. 
The  INTEGRAL \citep{li12} and BAT \citep{atel3936} 
flares at phase 0.55 and 0.57 respectively, together with the  AGL 
J1037-5708 flare \citep{atel3059} at phase 0.22, are marked with arrows.
}      
         
                \label{period} 
        \end{center}
        \end{figure}

	\section{Data reduction\label{data}}

The survey data collected with BAT between November 2004 and March 2013 were
retrieved from the HEASARC public 
archive\footnote{http://heasarc.gsfc.nasa.gov/docs/archive.html} and 
processed using a software dedicated to the analysis of the data from coded mask
telescopes \citep{segreto10}.
The source is detected with a significance of 26.8 standard deviations in the
15-60 keV energy band, where its signal is optimized: Fig.~\ref{map} (left panel) 
shows the significance map of the sky region around 4U~1036--56 in this energy 
band. The significance in the 15-150 keV band is 23.9 standard deviations.
The 15-60 keV light curve was extracted in 49493 bins with a variable time length 
ranging between 50 s and 6317 s and a constant pointing direction within each bin.
The central time of each bin was corrected to the Solar
System barycentre (SSB) using the task 
{\sc EARTH2SUN}
and the JPL DE-200 ephemeris \citep{standish82}.
 The background subtracted spectrum was 
extracted in eight energy channels and analyzed using the BAT redistribution 
matrix available in the Swift calibration 
database\footnote{http://swift.gsfc.nasa.gov/docs/heasarc/caldb/swift/}.

4U~1036--56 was observed once (ObsID 00032288001) with Swift-XRT on 
2012 February 17 for 3 ksec in
photon counting observing mode \citep{hill04}. The data were processed using the 
{\sc ftools} package with 
standard procedures ({\sc xrtpipeline} v.0.12.4), filtering and screening 
criteria, with standard grade filtering 0-12.

A first inspection of the light curve shows high variability, with a
maximum count rate of 1.2 c/s, indicating the need for a pile-up 
correction\footnote{see e.g. http://www.swift.ac.uk/analysis/xrt/pileup.php}.
Selecting only the time intervals where the source intensity exceeds 0.5 c/s, 
we have  compared the source
Point Spread function (PSF) with the expected PSF shape \citep{moretti05},
finding that a circular region with {\bf 4 pixels radius} (centered at the 
source position) must be excluded to avoid pile-up. 
Therefore, the source spectrum and light curve were extracted from an annular 
region of {\bf 4 and 30 pixels internal and esternal radii} respectively, centered 
on the source centroid as determined
with the task {\sc xrtcentroid} (RA$_{J2000}$=10h 37m 35.25s, Dec$_{J2000}$= -56$^{\circ}$ 47' 56.3'', with 3.6'' 
uncertainty at 90\% confidence level). 
The background {\bf for the lightcurve and spectral analysis} was extracted from an annular region  
with inner and outer radii of 40 and 70 pixels, respectively. 
This annular region is far enough to 
avoid the contamination from the PSF wings of 4U~1036--56 and does not contain any 
field source. {\bf  The source event arrival times were converted to the Solar
System barycenter using the task {\sc
barycorr};
the lightcurve was corrected for PSF and vignetting using the task {\sc
xrtlccorr}}.
The XRT ancillary response file generated with
{\sc xrtmkarf} 
accounts for PSF and vignetting correction;
we used the spectral redistribution matrix v013 available in the Swift calibration database. 
The spectral analysis was performed using {\sc xspec} v.12.5, after grouping 
the spectrum with a minimum of 20 counts per channel to allow the use of $\chi^2$
statistics.

We note that the X--ray coordinates of 4U 1036$-$56 determined 
with XRT are formally inconsistent with those (RA$_{J2000}$ = 10h 37m 35.5s,
Dec$_{J2000}$ = -56$^{\circ}$ 48' 11'') of the optical counterpart LS 1698 \citep{stephenson71} 
which are available 
e.g. from the CDS\footnote{http://cdsportal.u-strasbg.fr} Portal, 
the star being about 20$''$ from the X--ray centroid, thus well 
outside the XRT error circle.
However, an inspection of this sky region in the DSS-II-Red 
Sky Survey\footnote{http://archive.eso.org/dss/dss} shows that
no object is present at the CDS coordinates of LS 1698. 
Moreover, a bright optical object is instead present in the
DSS within the XRT error circle, and this is the same object
proposed by \citet[ see their Fig. 14]{motch97} as the optical 
counterpart of 4U 1036$-$56. 
Regrettably, \citet{motch97} did not report precise coordinates 
for LS 1698; we thus complete this information here by extracting 
it from the 2MASS archive, which has a precision of 0$\farcs$1 on 
both RA and Dec \citep{2mass}. According to this archive, 
the coordinates (J2000) of LS 1698 are RA$_{J2000}$ = 10h 37m 35.32s, Dec$_{J2000}$ = -56 $^{\circ}$ 47' 
55.8', at 0$\farcs$7 from the XRT error circle centroid, thus well 
within it.
Figure~\ref{map} (right panel) shows the XRT image of 4U~1036--56, with the 
95\% BAT error circle of 1.16' superimposed, and the position of the 
optical counterpart.

        \section{Timing analysis\label{timing}}

\begin{figure*}
\begin{center}
\centerline{\includegraphics[width=18cm]{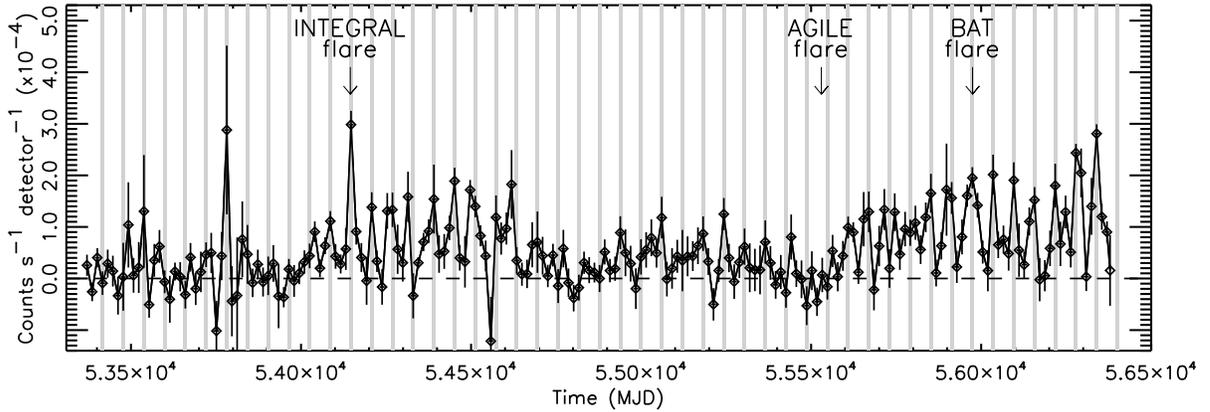}}
\caption[4U1036-56 BAT light curve]{15--60 keV BAT light
curve.  Each bin represents a time interval of  P$_0/4=15.25 $ d. 
The vertical shaded bars correspond to phase 0.375--0.6875 in
Figure~\ref{period}c. The epochs of the INTEGRAL \citep{li12} and BAT \citep{atel3936} 
flares, together with the epoch of the AGL J1037-5708 flare \citep{atel3059}
are marked with arrows.
                }
                \label{lc} 
        \end{center}
        \end{figure*}

We performed a timing analysis searching for long term periodic modulations 
in the 15--60 keV BAT  light curve of 4U~1036--56. We applied 
a folding algorithm to the  baricentered arrival times  searching
in the {\bf 0.5--1000\,days} time range with a step of P$^{2}/(N \,\Delta T)$,
where P is the trial period, $N=16$ is the number of  profile phase bins 
and {\bf $\Delta T=$262 Ms is the data time span}. 
The average rate in each phase bin  was evaluated by weighting the rates 
by the inverse square of the corresponding statistical errors  (see also
\citealp{cusumano10}). 
This is appropriate when dealing with a large span of rate errors 
and/or with background dominated data such as those from coded mask telescopes.\\
Figure~\ref{period} (a) shows the periodogram with several features emerging. 
A very prominent one, with a $\chi^2$ value of $\sim323$, is at
P$_0=61.0\pm0.2$ d,  where the period and its error $\rm \Delta P_0$ are  the
centroid of the peak and the standard deviation obtained from a Gaussian fit to
the $\chi^2$ feature at P$_0$; other significant features at higher periods 
correspond to multiples of P$_0$ (2P$_0$ to 10P$_0$). \\
The long term variability of the source causes the distribution of  $\chi^2$ in
Figure~\ref{period} (a) to deviate strongly both in average and in fluctuation amplitude
from the behavior expected for a white noise (average $\chi^2=N-1$) where the signal 
is dominated by statistical variations.
As a consequence the $\chi^2$ statistics cannot be applied and the significance for the 
presence of a feature has to be evaluated with different methodologies.  
For this purpose we applied the procedure described below in a few steps: \\ 
(1) - The significance of any feature needs to be evaluated with 
respect to the average level of the periodogram. Thus we fit the periodogram 
with a second order polynomial that describes
the trend of the  $\chi^2$ values in the periodogram. The best fit function was
then subtracted from the $\chi^2$ to obtain a flattened periodogram (hereafter,
z). The value of  z corresponding to P$_0$ is $\sim 279$.\\
(2) - Figure~\ref{period} (b) shows the histogram of the z values in the 
period range between 20 and 100 days (where the periodogram is
characterized by a noise level quite consistent with the noise level at 
P$_0$), excluding the values in the interval 
$\pm 3\Delta P_0$ around P$_0$. Its positive tail (beyond z=10) can be modeled 
with an exponential function, and its integral $\Sigma$
between 279 and infinity (normalized to the total area below the
distribution) represents the probability of chance occurrence of a z
equal to
or larger than  279. The area below the histogram was evaluated summing
the
contribution of each single bin from its left boundary up to  $z$=10 and
integrating the best-fit exponential function beyond $z$=10.
The probability of chance occurrence for $z\ge 279$
is $\Sigma=1.7 \times 10^{-11}$ and corresponds to $\sim6$ Gaussian
standard deviations for the significance of the feature at P$_0$.
The light curve profile (Fig.~\ref{period}, c) folded at P$_0$ with 
T$_{\rm epoch}$= 54841.199 MJD, is characterized by a roughly squared
shaped peak over a low intensity plateau.\\
{\bf Figure~\ref{lc} shows the 15--60 keV BAT light curve of 4U~1036--56 
sampled with a bin time of P$_0/4=15.25$ d. The vertical shaded area are in
phase with the peak in Figure~\ref{period}c (phase interval 0.375--0.6875).  
The light curve shows the presence of many flux enhancement, most of which
are coincident with the shaded bars.}\\
The \sw-XRT observation was performed as a follow up of the BAT flare 
\citep{atel3936} and it is at orbital phase 0.58 in
Fig.~\ref{period}(c). It is composed of two snapshots lasting $\sim$2200 s and 
800 s respectively, with an average pile-up corrected count rate of 0.6 
count s$^{-1}$. A light curve binned 
at intervals of 50 s (Fig.~\ref{xrt}) shows a variability within a factor of
10, with a modulation ($\sim$3 cycles) roughly consistent with the periodicity reported 
by \citet{reig99}. 

\begin{figure}
\begin{center}
\centerline{\includegraphics[width=7cm,angle=0]{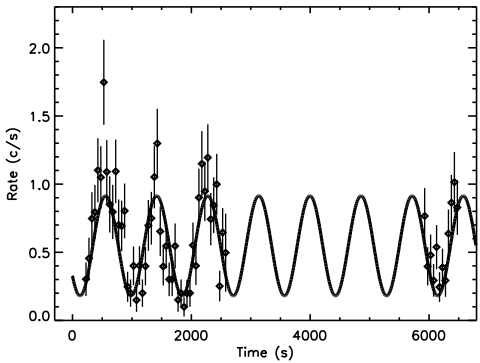}}
\caption[4U1036-56 XRT light curve]{0.2-10 keV XRT light
curve.  Each bin represents a time interval of 50 s. The solid line represents
a sinusoidal modulation with the pulse period of 860 s reported by \citet{reig99}.
                }
                \label{xrt} 
        \end{center}
        \end{figure}

\section{Spectral analysis\label{xrt}}

Since the XRT observation was performed during a flare, which 
in turn corresponds to the phase interval where several other intensity 
enhancements have been observed (see Figure~\ref{lc}), the broad band spectral 
analysis was performed coupling the soft X-ray spectrum and the 
Swift-BAT hard X-ray spectrum  selected at this phase interval (0.375--0.6875).

\begin{figure}
\begin{center}
\centerline{\includegraphics[width=5.0cm,angle=270]{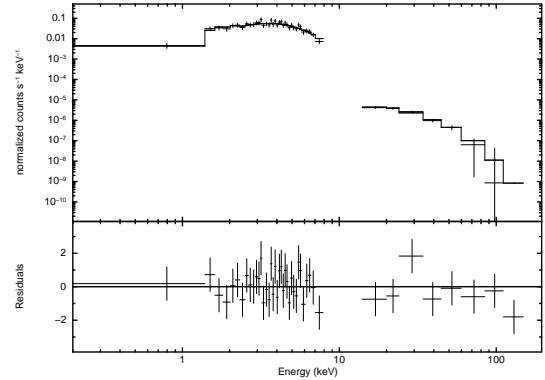}}
\caption[]{ 
4U1036-56  broad band (0.2--150 keV) spectrum. {\bf Top panel}: 
XRT and BAT data and best fit {\tt phabs*(cutoffpl)} model.
{\bf Bottom panel}: Residuals in unit of standard deviations.
}
                \label{spec}
        \end{center}
        \end{figure}

The broad band 0.2--150 keV spectral analysis was performed
introducing a multiplicative constant ($C_{BAT}$) in the model to
take into account both for the intercalibration systematics in the XRT and BAT 
responses and for the non simultaneity of the data.
We tried to fit the spectra with an absorbed power law model {\tt phabs*(powerlaw)}
that resulted unsatisfactory to describe the
data ($\chi^2$=91.2, 41 degrees of freedom [d.o.f.]). The fit residuals 
clearly indicated a significant steepening in the BAT energy range. 
The spectrum indeed results  well fitted  adopting the model {\tt phabs*(cutoffpl)}
with  a photon index $\Gamma\sim 0.6$ and a cutoff energy $\rm E_{cut}\sim16~keV$, 
with a  $\chi^2=32.5$, for 40 dof. The F-test for one
additional parameter yields a probability of chance improvement of $1.7\times
10^{-10}$ with respect to the absorbed powerlaw model. {\bf We observe that the
cutoff energy is close to the gap between the XRT and the BAT energy bands;
therefore its position could be affected by residual intercalibration
systematics between the two instruments.}
Figure~\ref{spec} (bottom panel) shows no significant residual pattern 
between data and model.
Table~\ref{fit} reports the best fit parameters (quoted errors are given at
90\% confidence level for a single parameter). {\bf The multiplicative constant
($C_{BAT}$) is marginally consistent with the ratio between the average count 
rate in the BAT spectrum and the count rate observed by BAT 
at the time of the XRT observation.}

\section{Discussion\label{discuss}}

We investigated the timing and spectral properties of the HMXB 
4U~1036--56 exploiting the Swift data recorded by the BAT and the XRT
telescopes. 
The timing analysis on the 100-month BAT survey light curve unveiled a periodic
modulation with a period of P$_0= 61.0\pm 0.2$ days. Such long periodicities are
typically associated with the orbit of binary systems. The profile of the
light curve folded at P$_0$ shows a roughly rectangular shape over 
a plateau of low intensity emission.

\begin{figure}
\begin{center}
\centerline{\includegraphics[width=7cm,angle=0]{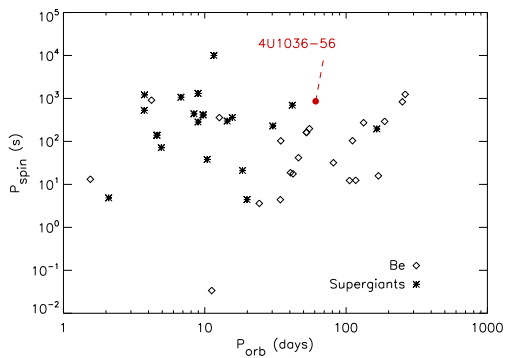}}
\caption[Corbet diagram]{
Corbet diagram for HMXBs with known spin and orbital period. Diamond and star points
represent the Be and supergiant systems, respectively. The red filled circle
marks the position of 4U~1036-56.
                }
                \label{corbet}
        \end{center}
        \end{figure}

The source can be located on the Corbet diagram \citep{corbet86}, as 
both its spin and orbital periods (P$_{spin}$, P$_{orb}$) are known. 
Figure~\ref{corbet} reports the pairs of P$_{spin}$ and 
P$_{orb}$ for the HMXBs listed in \citet{liu06},  
and for the HMXBs discovered more recently by 
INTEGRAL\footnote{http://irfu.cea.fr/Sap/IGR-Sources/}. 
4U~1036--56  lays at the boundary of the Be transients region, in agreement with
the classification of its companion star.

\begin{table}
\begin{tabular}{ r l l}
\hline
Parameter         & Best fit value & Units    \\ \hline \hline
 N$_{\textrm H}$  & $2.2^{+0.8}_{-0.6} \times 10^{22}$ & cm$^{-2}$\\
$\Gamma$          &$0.6^{+0.3}_{-0.3}$&          \\
$\rm E_{cut}$     &$16^{+5}_{-3}$ & keV\\
$N$               &$ 4.3^{+2.9}_{-1.6} \times 10^{-3}$ &ph $\rm keV^{-1} cm^{-2} s^{-1}$ at 1 keV  \\
$\rm C_{BAT}$     & $0.18^{+0.08}_{-0.05}$&\\
$F_{0.2-10~keV}$  &$ 8.7^{+0.7}_{-0.7} \times 10^{-11}$ & $\rm erg~s^{-1}~cm^{-2}$\\
$F_{15-150~keV}$  &$ 3.2^{+0.2}_{-0.3} \times 10^{-11}$ & $\rm erg~s^{-1}~cm^{-2}$ \\
$\chi^2$          &32.5 (40 dof) & \\ \hline
\end{tabular}
\caption{Best fit spectral parameters. The BAT spectra is selected in the phase
interval 0.375-0.6875 in Fig.~\ref{period} (c). We report
unabsorbed fluxes for the characteristic XRT (0.2--10 keV) and BAT (15--150 keV)
energy bands. $\rm C_{BAT}$ is the constant factor to be multiplied to the
model in the BAT energy range in order to match the BAT data.\label{fit}
}
\end{table}

Knowing the orbital period, we can use the third Kepler's law to derive the semi-major 
axis of the binary system. Assuming that $M_{\rm X}=1.4~M_{\odot}$ is the mass of the neutron star
and $M_{\star} = 17.5-20~M_{\odot} $ is the mass range for the spectral type of the companion star \citep{lang92}, we have:
\begin{equation}
a=(G P_0^2~(M_{\star}+M_{\rm X})/4\pi^2)^{1/3} \simeq 174-181 R_{\odot} 
\end{equation}
Considering that the radius of the companion star is  $R_{\star}=7.4-15~R_{\odot} $ 
\citep{lang92}, {\bf the semi-major
axis length corresponds to $ \sim 12-23.5~R_{\star}$. }
This wide orbital separation is  common among HMXB with Be star as companion.

We analyzed the broad-band (0.2--150 keV) spectrum of 4U~1036--56 using the 
XRT pointed observation data in the soft X-ray band and the BAT survey data 
selected in the 0.375--0.6875 phase interval. The spectrum is well modeled with a 
flat ($\Gamma\sim 0.6$) absorbed powerlaw with a cutoff at $\sim 16$ keV. 
The column density is $\rm \sim 2.2 \times 10^{22}~cm^{-2}$, which is a factor 
of $\sim 2$ larger than the maximum Galactic value in the direction of the source 
($\rm 9.1 \times 10^{21} cm^{-2}$  \citealp{dickey90}). This suggests an 
additional intrinsic absorption in the environment of the binary system.

\citet{li12}  discuss the possibility that 4U~1036--56 is associated to the 
unidentified source AGL~J1037-5708, that was detected in November 2012 during an
outburst above 100 MeV \citep{atel3059}. The association was initially 
suggested because of the spatial coincidence of the two sources.
Figure~\ref{lc} shows the epoch of the INTEGRAL \citep{li12} and 
BAT \citep{atel3936} flares, together with the epoch of the AGL J1037-5708 
flare \citep{atel3059}. A simultaneous flare in the MeV and in the hard X-ray energy ranges 
would be a strong evidence that the two sources are indeed associated. However, 
while the presence of {\bf many peaks} can be observed in the BAT light curve, 
the epoch of the MeV burst does not correspond to any significant intensity 
enhancement. Moreover we observe
that while the BAT and the INTEGRAL flare happen at a phase consistent
with the maximum of the folded profile (Figure~\ref{period}c), the MeV flare 
is at a phase of minimum intensity. On the other hand, 
it is not ensured that a MeV flare would automatically corresponds  to
a hard X-ray flare.


\section*{Acknowledgments}
This work has been supported
by ASI grant I/011/07/0.

\bsp

\label{lastpage}

\end{document}